# Goldbach Ellipse Sequences for Cryptographic Applications

Krishnama Raju Kanchu and Subhash Kak

***Abstract:*** The paper studies cryptographically useful properties of the sequence of the sizes of Goldbach ellipses. We show that binary subsequences based on this sequence have useful properties. They can be used to generate keys and to provide an index-based mapping to numbers. The paper also presents a protocol for secure session keys that is based on Goldbach partitions.

## I INTRODUCTION

The randomness properties of Goldbach sequences were presented in a recent paper [1]. These number-theoretic random sequences may be viewed from a computational complexity perspective [2] and they represent a class different from physics-based random sequences [3],[4]. The Goldbach partitions of an even number *n* are the ways in which the number is represented as a sum of two primes. The counts of these partitions vary from number to number. Thus 10 has two partitions 3+7 and 5+5 while 34 has four partitions 3+31, 5+29, 11+23, and 17+17. The Goldbach circle sequence of a given radius is the set of partitions from the even numbers of that radius. If the radius were 3, the circle sequence is:

    8: (5,11)

    10: (7,13)

    14: (11,17)

    …

Similarly, an ellipse can be constructed around an even number n, on the number line where the distance of the two extreme points from *n* is *j* and *k,* respectively. This will in general be represented by *(j,k)* ellipse.

For simplicity, we will now consider *j* to be 1; or, in other words, we consider *(1,k)* ellipse. The Goldbach ellipse for the point 2*n* is associated with the primes at (2n-m) and (2n +km). This leads to the definition of the corresponding m-sequence. This m-sequence has cryptographic applications.

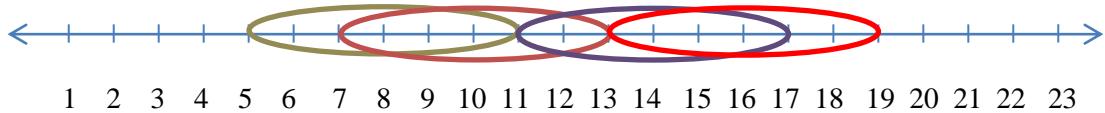

Figure1. The Goldbach ellipse of numbers 6,8,12 and 14 for k=5

The ellipse that is formed around an even number can differ from its distance from the even number depending on the values of k. The values of k are always an odd number greater than 1. Hence, ellipse of the form as shown in Figure 1 are obtained for k=5.

Table1. Goldbach ellipse partition sequence for k=5

| 2n | 2n-m | 2n+km | m | 4n+(k-1)m |
|---|---|---|---|---|
| 6 | 5 | 11 | 1 | 16 |
| 8 | 7 | 13 | 1 | 20 |
| 12 | 11 | 17 | 1 | 28 |
| 14 | 13 | 19 | 1 | 32 |
| 18 | 17 | 23 | 1 | 40 |
| 22 | 19 | 37 | 3 | 56 |
| 24 | 23 | 29 | 1 | 52 |
| 26 | 23 | 41 | 3 | 64 |
| 32 | 31 | 37 | 1 | 68 |
| 34 | 29 | 59 | 5 | 88 |
| 36 | 31 | 61 | 5 | 92 |
| 38 | 37 | 43 | 1 | 80 |
| 42 | 41 | 47 | 1 | 88 |

The m sequence generated for k=5 is shown in the fourth column of Table 1. This sequence will be different for different values of k. Another important point is the absence of the integral multiples of k in the 2n sequence in the above table. In other words, there are no ellipses that are present for even numbers that are integer multiples of k (i.e. no values of 2n that are multiples of 5). Hence we get the m sequence as 1,1,1,1,1,3,1,3,1,5,5,1,1.

The m-sequence can be converted into a binary sequence using the mapping: if $m$ mod 4 is 1 keep it as 1, if it is 3, keep it as -1. Thus the above m-sequence becomes the binary b-sequence

1 1 1 1 1 -1 1 -1 1 1 1 1 1 ….

The randomness properties of this binary sequence are presented in terms of its autocorrelation function.

## II AUTOCORRELATION FUNCTION

The autocorrelation function describes the relation between the two sequences. It measures the randomness of the sequences and its correlation with itself. If the function is close to being two-valued then it may be taken to be random. The autocorrelation function is mathematically given by

$$C(i) = 1/n \sum_{m=0}^{n-1} a_m a_{m+i}$$

From Figure 2 we see that C(i) characterizes a sequences that is highly random since it is 1 only for i=0 and it is close to zero for other values.

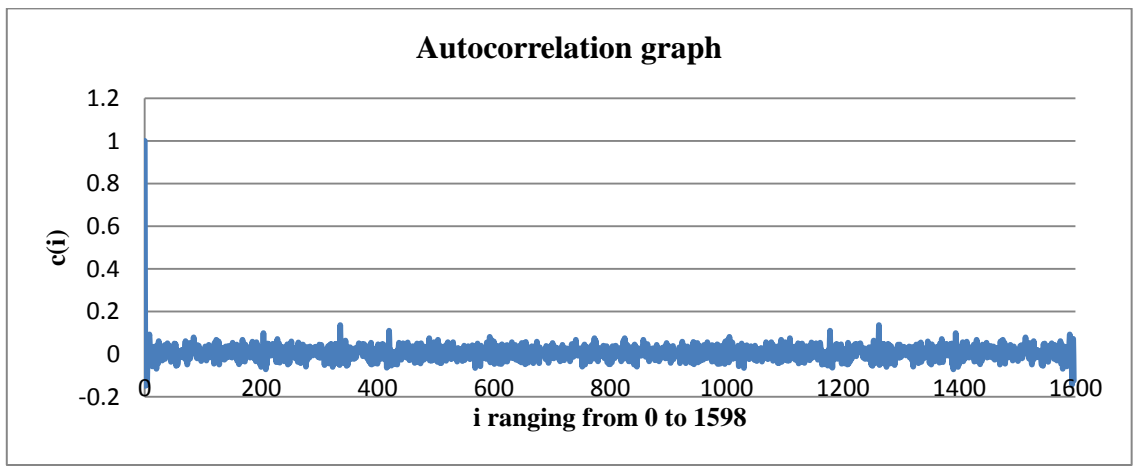

Figure 2. Autocorrelation graph for k=5.

An important problem is how to determine the m-sequence given a subsequence of the corresponding binary b-sequence. If the length of the substring is small it is likely to occur at many places in the sequence provided. On the other hand, if it is sufficiently long, it will be unique and its location within the original sequence can be found by inspection. If only the substring is transmitted the task of the eavesdropper to determine its place in the larger sequence can become a computationally hard puzzle.

# III SUBSTRING GENERATION

The m sequence is in terms of odd numbers for obvious reasons. These odd numbers are divided into two groups by taking modulus 4. We establish the count of variable length strings from the working set by using a moving window pattern. This leads to produce the count of different string patterns of the same length as well as different length. As shown below.

Table 2. Substring counter table for n < 2000, k=5

| Sub String | Count |
| --- | --- |
| 11 | 282 |
| 10 | 449 |
| 00 | 448 |
| 01 | 201 |
| 110 | 206 |
| 100 | 165 |
| 001 | 165 |

Also, the count of fixed length string is different for different lengths since, the count of the string depends on its occurrences in the main working set. Hence the count table is as shown below.

Table 3: String length and its count for n < 2000, k=5

| String Length | Count |
| --- | --- |
| 2 | 1380 |
| 3 | 1360 |
| 4 | 1420 |
| 5 | 1476 |
| 6 | 1502 |
| 7 | 1542 |
| 8 | 1561 |
| 9 | 1555 |
| 10 | 1567 |
| 11 | 1576 |
| 12 | 1584 |
| 13 | 1585 |
| 14 | 1584 |
| 15 | 1583 |
| 16 | 1585 |
| 17 | 1581 |
| 18 | 1580 |
| 19 | 1579 |
| 20 | 1578 |

The graph for the strings of different lengths is given in Figure 3.

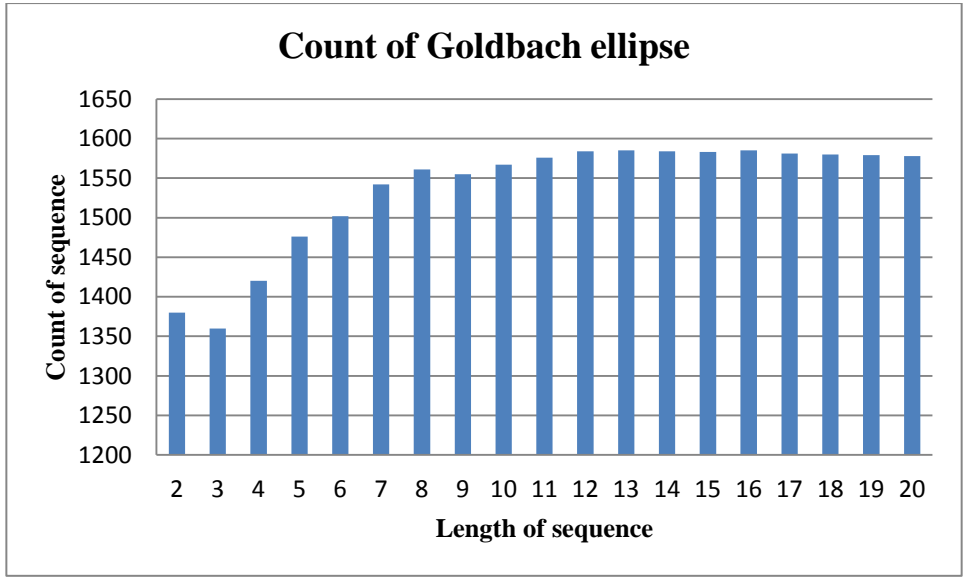

Figure3: Length of the string and its count for n < 2000, (k=5)

It is observed that even the working string set that is derived from the Goldbach conjecture by considering the even numbers with even number of partitions as 0 and odd number of partitions as 1 leads to the generation of the sequence 1110100001110100001010101111001100101...

As an ellipse could be generated from the desired values of k, similarly a circle, which is a special case of an ellipse, can be drawn by considering the value of k=1. We are interested in the properties of a circle as a special case of an ellipse to see if there is any structural difference in their behaviors and Figure 4 shows that the count of substrings is flatter for the circle.

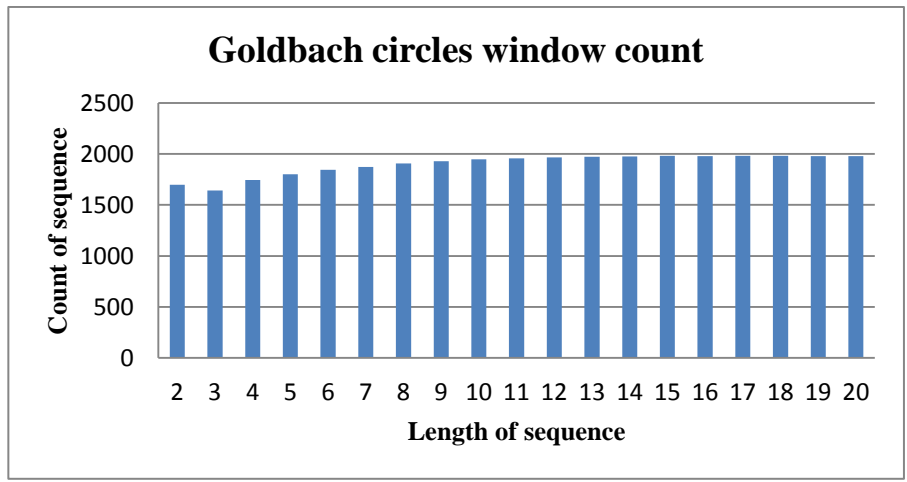

Figure 4. Length of string and count for n < 2000, k=1

In a circle, an even number is bounded by two primes on either side with an equal distance. And when the substrings of different length are examined we get Table 4.

Table 4. String length and its count for n < 2000 (k=1)

| String Length | Count |
|---|---|
| 2 | 1698 |
| 3 | 1642 |
| 4 | 1744 |
| 5 | 1801 |
| 6 | 1845 |
| 7 | 1871 |
| 8 | 1908 |
| 9 | 1927 |
| 10 | 1946 |
| 11 | 1957 |
| 12 | 1967 |
| 13 | 1972 |
| 14 | 1976 |
| 15 | 1980 |
| 16 | 1979 |
| 17 | 1981 |
| 18 | 1981 |
| 19 | 1979 |
| 20 | 1978 |

The results when shown on a graph produce the following pattern.

## IV A SECURE COMMUNICATION PROTOCOL

We now present a protocol to establish a secure link between two communicating parties Alice (A) and Bob (B). We assume that the network consists of a Certification Authority (CA) who mediates initial communication between the parties that leads to the choice of a session key. It is assumed that every party in the network has a secret key which is known only to the Certification Authority. How these secret keys are shared with CA will not be discussed in this paper.

Let the secret primes of A and B be $a$ and $b$, respectively. Knowing the identity of the parties, CA computes $a + b = n$ and then picks some other choice of partition so that $p + q = n$. The prime number $p$ is sent to both A and B in an encrypted form in terms of

Alice receives:   $p \oplus h(a)$

Bob receives:     $p \oplus h(b)$

where $h(.)$ is a standard hashing function and $\oplus$ is the mod 2 addition operation for the symbols in the binary representation of the numbers . Neither Alice nor Bob can compute $p$ unless they possess the secret numbers $a$ and $b$, respectively. This ensures that the communication between the parties is authenticated. Someone masquerading as Alice or Bob will not be able to extract the session key $p$. The secret prime $q$, available in the audit files of CA, will verify that the prime $p$ was generated based on knowledge of $a$ and $b$.

Once $p$ has been determined by Alice and Bob, they can use it directly as seed for a random number generator or they could use it to generate a pseudo-random d-sequence [6],[7],[8] that is added to the signals between the two parties.

The protocol is pictorially represented in Figure 5.

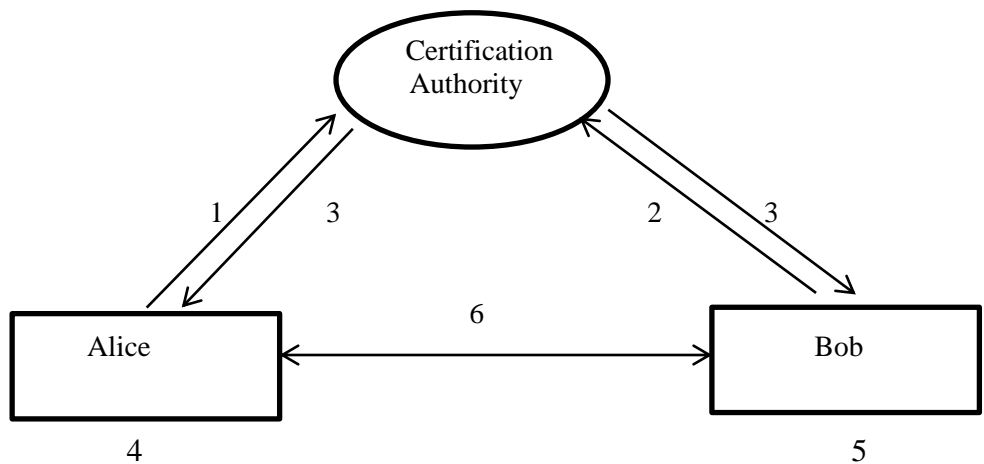

Figure 5. Protocol to establish session keys

Step1. Alice (A) informs the Certification Authority (CA) of her wish for a secure session key for communication with Bob (B)

Step2. Bob sends agreement to CA for such a communication.

Step3. CA uses the stored $a$ and $b$ values to compute $h(a)$ and $h(b)$. CA chooses $p+q = a+b$, where $p$ is the session key and $q$ is the audit key. CA sends $p \oplus h(a)$ to A and $p \oplus h(b)$ to B.

Step4. Alice computes the value $p$ using its secret $a$.

Step5. Bob computes the value $p$ using its secret $b$.

Step6. Alice and Bob start communicating with each other using p as seed for RNG.

This protocol can be made stronger by the use of additional random numbers so that replay attacks cannot be mounted.

## V CONCLUSIONS

This paper has presented further properties of Goldbach ellipse sequences that are useful in cryptographic applications. In particular, the subsequences associated with Goldbach ellipses were discussed. These subsequences can be used to index random keys.

The paper also presents a protocol for secure session keys that is based on Goldbach partitions. The fact that a number can be partitioned in a variety of ways provides properties that can be conveniently exploited in the design of security protocols.

**Acknowledgement.** This research was supported in part by research grant #1117068 from the National Science Foundation.